\documentclass[conference]{IEEEtran}
\usepackage[utf8]{inputenc}

\usepackage{cite}
\usepackage{siunitx}
\usepackage{pbox}
\usepackage{rotating}
\usepackage{tabularx}
\usepackage{afterpage}
\usepackage{amsmath,amssymb,amsfonts,bm}

\usepackage{graphicx}
\usepackage{float}
\usepackage{mathtools, cuted}
\usepackage{textcomp}
\def\BibTeX{{\rm B\kern-.05em{\sc i\kern-.025em b}\kern-.08em
T\kern-.1667em\lower.7ex\hbox{E}\kern-.125emX}}
\usepackage{verbatim}
\usepackage[overload]{textcase}
\usepackage{verbatim}
\usepackage{gensymb}
\usepackage{algorithm}
\usepackage{algpseudocode}
\usepackage{physics}    %for derivation
\usepackage{xspace}
\usepackage{todonotes}
\usetikzlibrary{trees}

\usepackage{textcomp}
\usepackage{tikz} % the first step using the package tikz
\usetikzlibrary{backgrounds}
\usepackage{pgfplots} % powerful visualization tool and ideal for creating scientific/technical graphics.
\usepackage{pgfplots}
\usepgflibrary{plotmarks} 
\usepgfplotslibrary{colorbrewer}
\usepgfplotslibrary{external}
\usetikzlibrary{decorations.text}
\usepackage{etoolbox}
\usepackage{varwidth}
\usepackage{tikz-3dplot}
\pgfplotsset{compat=1.13}

\usepackage{pgfplots}

\graphicspath{ {images/} }

\newcommand{\nasrin}[1]{{\textcolor{black}{#1}}}
\newcommand{\bho}[1]{{\textcolor{black}{#1}}\xspace}

\makeatletter
\newcommand{\algmargin}{\the\ALG@thistlm}
\makeatother
\newlength{\whilewidth}
\algdef{SE}[parWHILE]{parWhile}{EndparWhile}[1]
  {\parbox[t]{\dimexpr\linewidth-\algmargin}{%
     \hangindent\whilewidth\strut\algorithmicwhile\ #1\ \algorithmicdo\strut}}{\algorithmicend\ \algorithmicwhile}%
\algnewcommand{\parState}[1]{\State%
  \parbox[t]{\dimexpr\linewidth-\algmargin}{\strut #1\strut}}

\title{On-Board Federated Learning for Dense LEO Constellations}

\author{\IEEEauthorblockN{Nasrin Razmi\IEEEauthorrefmark{1}\IEEEauthorrefmark{2}, Bho Matthiesen\IEEEauthorrefmark{1}\IEEEauthorrefmark{2}, Armin Dekorsy\IEEEauthorrefmark{1}\IEEEauthorrefmark{2}, and Petar Popovski\IEEEauthorrefmark{3}\IEEEauthorrefmark{1}}

\IEEEauthorblockA{\IEEEauthorrefmark{1}University of Bremen, U Bremen Excellence Chair, Dept. of Communications Engineering, Germany}

\IEEEauthorblockA{\IEEEauthorrefmark{2}Gauss-Olbers Center c/o University of Bremen, Dept. of Communications Engineering, Germany}

\IEEEauthorblockA{\IEEEauthorrefmark{3}Department of Electronic Systems, Aalborg University, Aalborg, Denmark}

{{Email:  \{razmi, matthiesen, dekorsy\}@ant.uni-bremen.de, petarp@es.aau.dk} \thanks{This work was funded in part by
by the German Research Foundation (DFG) under Germany's Excellence Strategy (EXC 2077 at University of Bremen, University Allowance).}}}

\begin{document}
%\bstctlcite{IEEEexample:BSTcontrol}

\maketitle
\begin{abstract}

Mega-constellations of small-size Low Earth Orbit~(LEO) satellites are currently planned and deployed by various private and public entities. While global connectivity is the main rationale, these constellations also offer the potential to gather immense amounts of data, e.g., for Earth observation.
Power and bandwidth constraints together with motives like privacy, limiting delay, or resiliency make it desirable to process this data directly within the constellation. We consider the implementation of on-board federated learning~(FL) orchestrated by an out-of-constellation parameter server~(PS) and propose a novel communication scheme tailored to support FL. It leverages intra-orbit inter-satellite links, the predictability of satellite movements and partial aggregating to massively reduce the training time and communication costs. In particular, for a constellation with 40 satellites equally distributed among five low Earth orbits and the PS in medium Earth orbit, we observe a 29$\bm{\times}$ speed-up in the training process time and a 8$\bm{\times}$ traffic reduction at the PS over the baseline.

%for the 98th percentile of test accuracy
%The canonical star architecture FL is not the best fit for the satellite scenarios because the location of satellites in orbits causes the parameter server~(PS) and satellites not to be visible to each other all the time. We address this issue and introduce a synchronous FL with star architecture of ring clusters for mega-constellations satellites with intra-plane inter-satellite links (ISLs), named FedISL. In FedISL, we use the predictability of the position of satellites and PS to schedule the transmissions among satellites and PS. We consider three different scenarios in which the PS is 1) a Medium Earth Orbit~(MEO) satellite, 2) ground station (GS) in the North Pole, and 3) GS placed in Bremen. We also examine the canonical FL, named FedNonISL, for the defined scenarios and compare the results with the FedISL. We show that by the FedISL scheme, we can have a more accurate model in a shorter time. Moreover, using MEO as the PS can provide a higher convergence rate for both algorithms.

%The results show that the FedISL scheme improves the test accuracy by 13\% and \%42 compared to the FedNonISL scheme for the PS as Medium Earth Orbit (MEO) and North Pole GS, respectively. Moreover, in the case of GS in Bremen, both schemes can provide low test accuracy.
\end{abstract}


\section{Introduction} 
%\textcolor{red}{cost of more GSs}

Constellations of low Earth orbit (LEO) satellites with small form factor are emerging to enable ubiquitous connectivity and support various Earth observation applications. Their key benefit over larger traditional satellites are vastly reduced development and deployment costs, together with lower latency in ground to satellite connections and denser Earth coverage \cite{sweeting2018modern,9217520}.
%\bho{ in higher orbits}
One major objectives for such constellations is Earth observation. Through Earth observation, valuable information can be provided for decision-makers in applications such as urban planning and disaster management \cite{mateo2021towards}. For example, the Disaster Monitoring Constellation~(DMC) \cite{da2005first} is one of the first-ever constellations aimed at disaster management. The declining trend in the cost of sending small satellites to space is likely to further increase this trend.

A challenge for these satellite constellations is that downlink transmission of huge volumes of collected data is unreasonable from an environmental perspective and leads to unnecessary spectrum utilization. The obvious solution to this is on-board processing of the data and then sending only relevant information to the Ground Station (GS).
%One major use case is data-driven machine learning \cite{vazquez2021machine,rolf2021generalizable,giuffrida2020cloudscout} based on federated learning (FL) \cite{mcmahan2017communication}, where the training of a machine learning model is implemented directly on the satellites without exchanging any raw data \cite{razmi2021ground}.

%We should mention that sats might belong to different sat operatores. These operators  perform joint earth observation using a ML-model they agreed on, but w/o the need to change their data measured dat. Armin comment
%,rolf2021generalizable
One major use case is data-driven machine learning \cite{vazquez2021machine,giuffrida2020cloudscout} based on federated learning (FL) \cite{mcmahan2017communication}, where the training of a machine learning model is implemented directly on the satellites without exchanging any raw data \cite{razmi2021ground}. \nasrin{FL can work perfectly in the constellations with satellites from different operators. These operators perform joint Earth observation using an agreed ML model, but without the need to exchange their raw data.}
In particular, this process is orchestrated by a parameter server (PS) that maintains a global version of the model and distributes it to the \nasrin{satellites}. These improve the model parameters using a local gradient descent (GD) procedure and then return the updated model parameters to the PS. After all updates are received, the PS aggregated them and starts a new round ("epoch").

In general, the satellites have intermittent but predictable connectivity to the PS due to their orbital movement \cite{razmi2021ground}.
We propose a communication scheme specifically tailored to running a synchronous FL procedure within a satellite mega-constellation orchestrated by an out-of-constellation PS, i.e., either a satellite in sufficiently different orbital height or a GS. This scheme leverages intra-plane ISLs to bridge non-availability periods between individual satellites and the PS. This is combined with partial aggregation to avoid unnecessary transmissions. \nasrin{We focus on enabling FL with only intra-plane ISLs rather than inter-plane ISLs, which connect satellites from different orbits. The reason is that inter-plane ISLs are highly dynamic, can be affected by Doppler shift, and require much more careful connection planning \cite{9327501}.}

A similar scenario is considered in \cite{razmi2021ground}. The main difference to this paper is that \nasrin{in \cite{razmi2021ground}} no ISLs are considered and each satellite needs to individually contact the PS. The underlying network topology is a star architecture of ring clusters as a result of having several circular orbits with one PS. Several previous works have considered this architecture \cite{lee2020tornadoaggregate,
%\bho{scenario}
duan2020self,wang2021efficient}. However, none addresses the intrinsic properties of satellite constellations such as the non-visibility and the predictability behavior.
%are not taken into consideration.
%%%%%%%%%%%%%%%%%%%%%%%%%%%%%%%%%%%%%%%%%%%%%%%%%%%%%%%%%%%%%%%%%%%%%%%%%%% 
\section{System Model and Problem Formulation}
Consider the satellite constellation illustrated in Fig.~\ref{fig:system model} consisting of $P$ orbital planes. Orbit $p$ contains $K_p$ equally spaced satellites with unique IDs $\{ k_{p,1}, \dots, k_{p,K_p} \} = \mathcal K_p$ such that $\mathcal K_{p_1} \cap \mathcal K_{p_2} = \emptyset$, for all $p_1 \neq p_2$, and $\mathcal K = \bigcup_{p = 1}^P \mathcal K_p = \{ 1, 2, \dots, K \}$, where $K = \sum_{p=1}^P K_p$ is the total number of satellites.
For orbit $p$, $h_p$ and $i_p$ denote the altitude and inclination, respectively. Each satellite in orbit $p$ moves with speed $v_p = \sqrt{\frac{\mu}{h_p+r_E}}\,\si{\meter/\second}$, where $r_E = 6371\,\si{\km}$ is the Earth radius and $\mu = 3.98 \times 10^{14}\,\si{\meter^3/\second^2}$ is the geocentric gravitational constant. The orbital period of the satellites in orbit $p$ is $T_p = \frac{2\pi(r_E + h_p)}{v_p}$. 

%%%%%%%%%%%%%^^^^^^^^^^^^^^^^^^^^^^^^^^^^^^
\begin{figure}
\centering
\tikzset{squares/.style={mark=square*, fill=pink, mark size=6pt}}

\tikzset{inter/.style={black, very thick}}
\tikzset{intra/.style={black, thick}}

\tdplotsetmaincoords{90}{0}
\tikzsetnextfilename{Walker_delta_diag}

\begin{tikzpicture}[tdplot_main_coords, scale=0.3]
\pgfmathsetmacro{\nop}{5}
\pgfmathsetmacro{\maxind}{\nop-1}
\pgfmathsetmacro{\eps}{360/\nop}
\pgfmathsetmacro{\r}{6.378}
\pgfmathsetmacro{\h}{\r+0.45}
\pgfmathsetmacro{\np}{3}
\pgfmathsetmacro{\delta}{180-53}
\pgfmathsetmacro{\pdelta}{90-53}

\pgfmathsetmacro{\h}{\r+0.55}
\foreach \n in {0}
{
\pgfmathsetmacro{\offset}{\n*360/\nop}
\tdplotsetrotatedcoords{140-\n*\eps}{\delta}{100-\offset}    %rotation (z,y,x)

\begin{scope}[tdplot_rotated_coords]
\draw[intra] (\r+0.5,0,0) arc (0:80:\r+0.5);
\draw[intra] (\r+0.5,0,0) arc (0:-154:\r+0.5);
%\draw[intra] (\r+0.5,0,0) arc (0:-180:\r+0.5);

\pgfmathsetmacro\k{(4-1)*20}
\foreach \ang in {0,1,...,\np}{
	\pgfmathsetmacro\ds{\ang+Mod(1,2)/2}
	\draw[squares] plot coordinates {($(0,0,0)+(180+\ds*180/\np:\r+0.5+1/100)$)};
}
\end{scope}
}

%%%%%%%%%%%%%%%%%%%%%%%%%%%%%%%%%%%%%%%%%%%%%%%%%

\pgfmathsetmacro{\h}{\r+0.55}

\pgfmathsetmacro{\offset}{0*360/\nop}
\tdplotsetrotatedcoords{40-0*\eps}{\delta}{-40-\offset}    %rotation (z,y,x)
\foreach \n in {1}
{
\begin{scope}[tdplot_rotated_coords]
%\draw[intra] (\r+0.5,0,0) arc (0:92:\r+0.5);
\draw[intra] (\r+0.5,0,0) arc (0:-140:\r+0.5);
\draw[intra] (\r+0.5,0,0) arc (0:95:\r+0.5);
%\draw[intra] (\r+0.5,0,0) arc (0:-60:\r+0.5);

\pgfmathsetmacro\k{(4-\n)*20}
\foreach \ang in {0,1,...,\np}{
	\pgfmathsetmacro\ds{\ang+Mod(\n,2)/2}
	\draw[squares] plot coordinates {($(0,0,0)+(190+\ds*180/\np:\r+0.5+\n/100)$)};
}
\end{scope}
}

%%%%%%%%%%%%%%%%%%%%%%%%%%%%%%%%%%%%%%%%%%%%
\pgfmathsetmacro{\h}{\r+0.55}
\foreach \n in {4}
{
\pgfmathsetmacro{\offset}{\n*360/\nop}
\tdplotsetrotatedcoords{90-\n*\eps}{\delta}{270-\offset}
\begin{scope}[tdplot_rotated_coords]
\draw[intra] (\r+0.5+\n/100,0,0) arc (0:210:\r+0.5+\n/100);
\draw[intra] (\r+0.5+\n/100,0,0) arc (0:-20:\r+0.5+\n/100);
%\draw[intra] (\r+0.5+\n/100,0,0) arc (0:-210:\r+0.5+\n/100);

\pgfmathsetmacro\k{(4-1)*20}
\foreach \ang in {0,1,...,\np}{
	\pgfmathsetmacro\ds{\ang+Mod(1,2)/2}
	\draw[squares] plot coordinates {($(0,0,0)+(\ds*180/\np:\r+0.5+1/100)$)};
}
\end{scope}
}
%%%%%%%%%%%%%%%%%%%%%%%%%%%%%%%%%%%%%%%%%%%%%%%%%

\pgfmathsetmacro{\h}{\r+0.55}
\foreach \n in {1}
{
\pgfmathsetmacro{\offset}{\n*360/\nop}
\tdplotsetrotatedcoords{90-\n*\eps}{\delta}{270-\offset}
\begin{scope}[tdplot_rotated_coords]
\draw[intra] (\r+0.5+\n/100,0,0) arc (0:190:\r+0.5+\n/100);
\draw[intra] (\r+0.5+\n/100,0,0) arc (0:-30:\r+0.5+\n/100);
\pgfmathsetmacro\k{(4-\n)*20}
\foreach \ang in {0,1,...,\np}{
	\pgfmathsetmacro\ds{\ang+Mod(\n,4)/4}
	\draw[squares] plot coordinates {($(0,0,0)+(\ds*180/\np:\r+0.5+\n/100)$)};
}
\end{scope}
}

%%%%%%%%%%%%%%%%%%%%%%%%%%%%%%%%%%%%%%%%%%%%%%%

\pgfmathsetmacro{\h}{\r+0.55}
\foreach \n in {1}
{
\pgfmathsetmacro{\offset}{\n*360/\nop}
\tdplotsetrotatedcoords{90}{\delta}{280-\offset}
\begin{scope}[tdplot_rotated_coords]
\draw[intra] (\r+0.5+\n/100,0,0) arc (0:100:\r+0.5+\n/100);
\draw[intra] (\r+0.5+\n/100,0,0) arc (0:-160:\r+0.5+\n/100);
%\pgfmathsetmacro\k{(4-\n)*20}
\foreach \ang in {0,1,...,5}{
	\pgfmathsetmacro\ds{\ang+Mod(\n,4)/4}
	\draw[squares] plot coordinates {($(0,0,0)+(200+\ds*240/5:\r+0.5+\n/100)$)};
}
\end{scope}
}

%%%%%%%%%%%%%%%%%%%%%%%%%%%%%%%%%%%%%%%%%%%%%%%%%
 % \shade[ball color = gray!40, opacity = 0.4] (0,0) circle (6.371cm);

% Earth in Background
\begin{pgfonlayer}{background}  
        \node {\includegraphics[scale=.20]{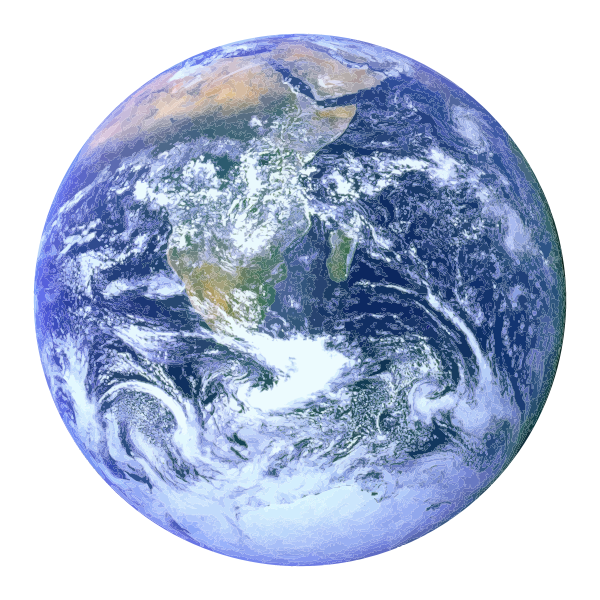}};  
\end{pgfonlayer} 
%%%%%%%%%%%%%%%%%%%%%%%%%%%%%%%%%%%%%%%%%%%%%%

\end{tikzpicture}
    \vspace{-2.5ex}
    \caption{Walker constellation with $K = 40$ satellites in $P = 5$ orbits. Each orbit consists of $K_p = 8$ satellites.}
    %located in altitude $h_p = \SI{2000}{\km}$ and moving with speed $v_p = \SI{6895}{\meter/\second}$. }
    \label{fig:system model}

\end{figure}
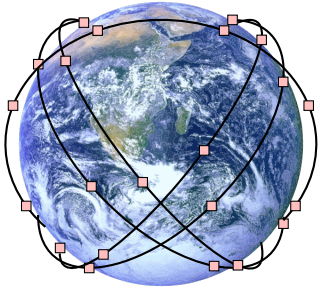
%%%%%%%^^^^^^^^^^^^^^^^^^^^^^^^^^^^^^^^^^^^^^^^

\iffalse
\begin{figure}
    \centering
    \includegraphics[scale = 0.50]{system_model1.png}
    \caption{Walker constellation with $K = 40$ satellites in $P = 5$ orbits. Each orbit consists of $K_p = 8$ satellites, located in altitude $h_p = \SI{2000}{\km}$ and moving with speed $v_p = \SI{6895}{\meter/\second}$. }
    \label{fig:system model}
\end{figure}
\fi
\subsection{Computation Model}
Each satellite within the constellation $\mathcal K$ has a local data set $\mathcal D_k$ that is used to train a ML model using the Federated Averaging (FedAvg) algorithm from \cite{mcmahan2017communication}. The training process is orchestrated by another satellite in a different orbit or by a GS, denoted as the PS. The local data sets are neither known nor shared with the PS or other satellites. The collaborative goal is \nasrin{to learn the model parameters that minimize a global loss function}
%The collaborative goal is to minimize a global loss function
%is to learn the parameter vector \omega that minimizes. Armin comment
\vspace{-2mm}
\begin{equation}
    F(\pmb{w}) = \sum\nolimits_{k\in\mathcal K} \frac{D_k}{D} F_k(\pmb{w}),
\end{equation}
where $D_k$ is the size of $\mathcal D_k$, $D = \sum_{k\in\mathcal K} D_k$ is the total number of training samples, $\pmb{w}$ is the parameter vector describing the model, and $F_k(\pmb{w})$ is the local loss function at satellite $k$. We consider a local loss function
\begin{equation} \label{eq: loss function of satellite k}
    F_k(\pmb{w}) = \frac{1}{D_k} \sum\nolimits_{\pmb{x}\in\mathcal D_k} f_k(\pmb{x}, \pmb{w}),
\end{equation}
\vspace{-0.5mm}
where $f_k(\pmb{x}, \pmb{w})$ is the per-sample loss function at satellite $k$ that depends on the concrete learning scenario at hand and is possibly nonconvex. We are interested in the case with full device participation \cite{li2020unified}.

The FedAvg algorithm is based on 1) transmitting the global parameters $\pmb{w}$ from the PS to the satellites, 2) training the model at the satellites using local GD, 3) sending the obtained local parameters to the PS by satellites, and 4) aggregating the parameters by the PS. This process continue for several epochs until convergence.
The local training steps are referred to ML steps in each client.
Every satellite $k$ minimizes its loss function using $I$ iterations of local GD and updates the local version of the global model parameters as
\begin{equation} \label{eq: GD}
    {\pmb{w}_k^{n,i}} = {\pmb{w}_k^{n, i-1}} -\eta\bigtriangledown{F_k({\pmb{w}_k^{n, i-1}})}, 
\end{equation}
\vspace{-0.5mm}
where $\eta$ is the learning rate, $n$ the global epoch, $i \ge 1$ the local iteration in epoch $n$, and $\pmb{w}_k^{n,0}$ = $\pmb{w}^n$ is initialized with the received global model parameters $\pmb{w}^n$. After receiving the local updates $\big(\pmb{w}_k^{n,I}\big)_{k\in\mathcal K}$ from all \nasrin{workers}, the PS aggregates them into a new version of the global model \nasrin{parameters} as
\begin{equation} \label{eq: FL parameters}
    {\pmb{w}^{n+1}} = \sum\nolimits_{k=1}^{K}  \frac{D_k}{D} \pmb{w}_k^{n, I}
\end{equation}
and advances into the next epoch until the global convergence criterion is met \cite{mcmahan2017communication}.

The time required for learning in every satellite $k$ is proportional to the
size of the data S($\mathcal D_k$) \nasrin{in bits} and the CPU frequency $\nu_k$ at satellite $k$, i.e.,
%\bho{number of samples $D_k$ and the CPU frequency $\nu_k$ at satellite $k$, i.e.,}
\vspace{-1.5mm}
\begin{equation} \label{eq: Time of learning}
    t_l(k) = \frac{c_k  S(\mathcal D_k)}{\nu_k},
\end{equation}
\vspace{-0.5mm}
where $c_k$ is the required number of CPU cycles to process a single data sample. This computation time model has been proposed for FL in \cite{nguyen2020toward}.

%\bho{where $c_k$ is the required number of CPU cycles to process a single data sample. This computation time model has been proposed for FL in \cite{nguyen2020toward}.}

\subsection{Communication Model}
Communication between satellites is feasible if the line of sight is unobstructed by Earth. By straightforward geometric considerations, it can be shown that this is the case if $d(k,i) < d_{Th}(k,i)$, where $d(k, i)$ is the Euclidean distance between satellites $k$ and $i$ and $d_{Th}(k, i)$ is a threshold value computed as
\begin{equation} \label{eq: visibility of satellites}
    d_{Th}(k,i) = \sqrt{{(h_{\pi(k)}+r_E)^2}-{r_E^2}}+\sqrt{{(h_{\pi(i)}+r_E)^2}-{r_E^2}}.
\end{equation} 
for the case that $k$ and $i$ are two satellites. $\pi(k)$ is the index of satellite $k$'s orbit. For the satellite-to-ground communications, based on the minimum elevation angle $\alpha_e$, the PS and satellite are visible if $\frac{\pi}{2} - \angle (\vec r_{PS}, \vec r_k - \vec r_{PS}) \ge \alpha_e$ where $\vec r_{k}$ and $\vec r_{PS}$ denote the position of satellite $k$ and PS.
 %The same condition also applies to satellite-to-ground communications.

Assuming Gaussian channels, the \bho{maximum achievable} data rate for the transmission between satellites $k$ and $i$ is 
\begin{equation} \label{eq: rate}
    r(k,i) = B \log_2 \left(1+\mathrm{SNR}(k,i)\right),
\end{equation}
where $B$ is the channel bandwidth, and $\rm{SNR}(k,i)$ is the signal to noise ratio (SNR)  which can be expressed as \cite{ippolito2017satellite, 9327501}
\begin{equation} \label{eq: SNR}
    \mathrm {SNR}(k,i) = \frac{P_t G_k(i) G_i(k) }{N_0  L(k,i)}
\end{equation}
%if $d(k, i) \ge d_\mathrm{Th}(k, i)$ and zero otherwise,
if $k$ and $i$ are visible and zero otherwise,
where $G_j(l)$ is the average antenna gain of satellite $j$ towards satellite $l$, $N_0 = k_B T B$ is the total noise power with $k_B = 1.380649 \times
%\bho{is the noise spectral density with}
10^{-23}\,\si{\joule/\kelvin}$ being the Boltzmann constant, $T$ is the receiver noise temperature. The free space path loss $L(k,i)$ is
% in kelvin
\begin{equation} \label{eq: LOSS}
    L(k,i) = \left(\frac{4\pi f_c d(k,i)}{c}\right)^2,
\end{equation}
\vspace{-0.5mm}
where $f_c$ and $c$ are the carrier frequency and the speed of light, respectively.
The time for transmitting a parameter vector $\pmb{w}$ from satellite $k$ to satellite $i$ is the sum of propagation, transmission, and communication processing delays, i.e.,
\begin{equation} \label{eq: Time for transmitting}
    t_c(k, i)= \underbrace{\frac{S(\pmb{w})}{r(k,i)}}_{\mathclap{\text{Transmission}}} + \underbrace{\frac{d(k,i)}{\mathstrut c}}_{\mathclap{\text{Propagation}}} + t_s(k) + t_r(i),
\end{equation}
where $t_s(k)$ and $t_r(k)$ are the processing delays in the transmitter and receiver of satellite $k$, respectively, and $S(\pmb{w})$ is the data size of $\pmb{w}$ in bits. Observe that the size of $\pmb{w}$ is model dependent and remains fixed during the whole training process. This model is also used for ground to satellite links if necessary. \bho{Observe that \eqref{eq: Time for transmitting} is not dependent on the usage of the Shannon rate in \eqref{eq: rate} but can also be any other rate function that adequately models the data rate over the considered link.}

\section{Satellite Communication for FL}

%synchronous FL implementation
Connections between the PS and \nasrin{workers} are the necessary condition to exchange the model parameters in canonical FedAvg \cite{mcmahan2017communication}. In the satellite scenarios, the location and motion of each LEO satellite and PS cause them to visit for a while. After the visiting duration, the connection between them is implausible due to the blockage by the Earth. This pattern of vising, followed by non-visibility periods continues. 

The duration and start of each visit do not follow a regular pattern because both satellites and PS move and they are most probably located in different altitudes. In rare cases, the visiting follows a regular pattern. Take the PS in the north pole and the satellites with inclination \ang{90} as an example. In this case, the duration between two visits lasts exactly one orbital period $T_p$ because the PS does not move. However, because the inclinations are not always exactly \ang{90} and the PS does not located in the North pole, the time between visits can be less or more than the $T_p$.

\begin{figure}
    \centering
    \includegraphics[scale=0.47]{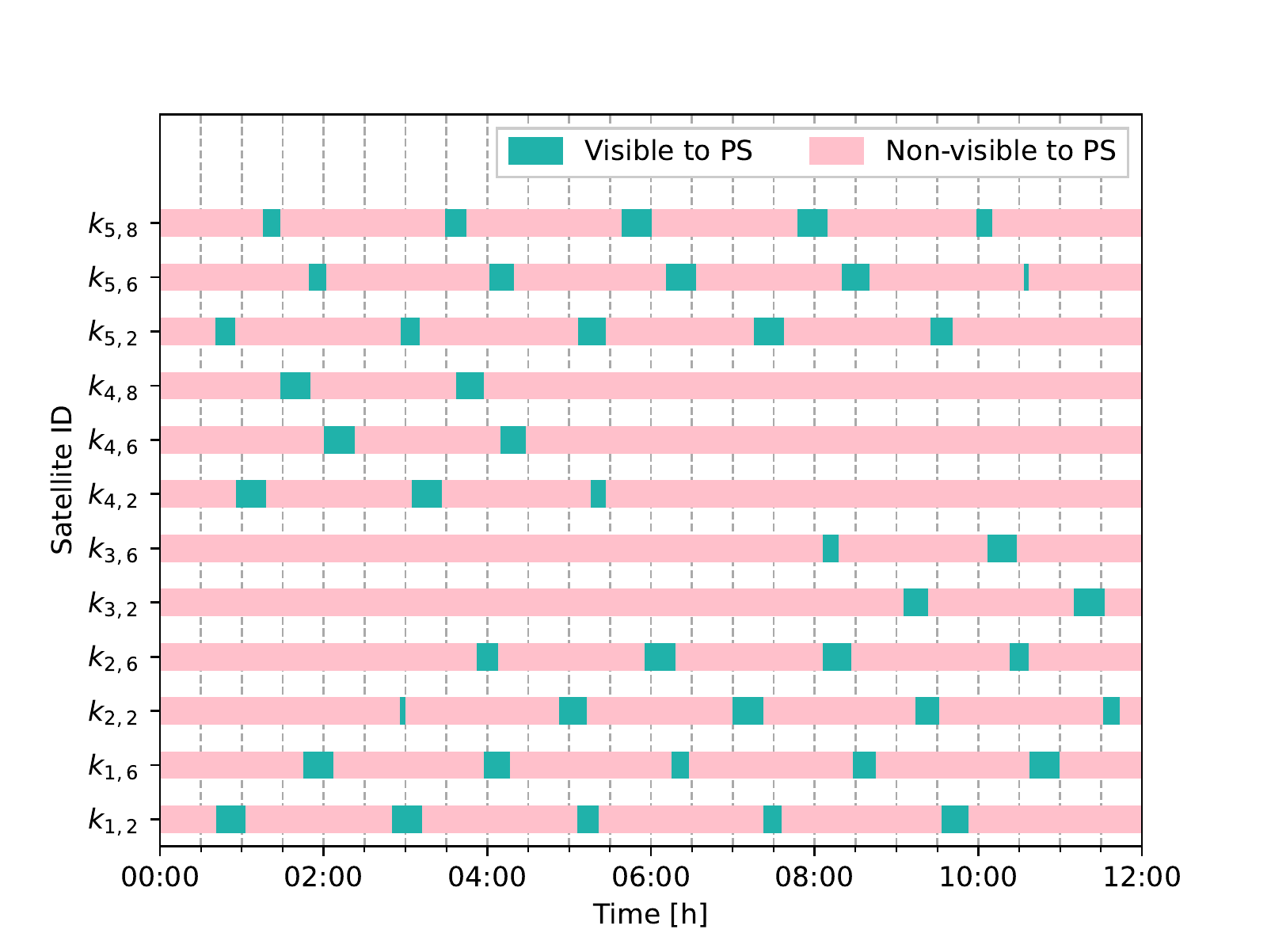}
    \vspace{-2.5ex}
    \caption{Visiting patterns between 12 satellites and a PS located in Bremen over 12 hours, where $k_{p,i}$ denotes the $i^{th}$ satellite in orbit $p$.}
    \label{fig:Visiting patterns}
\end{figure}

For more clarification, we have shown the visiting between satellites and the PS in the Bremen in 12 hours as Fig. \ref{fig:Visiting patterns} in five different orbits for the system model presented in Fig. \ref{fig:system model}. As this figure shows, satellites have different visiting patterns \nasrin{as a result of orbit specification, satellite and Earth movements}. Take the satellites with IDs $k_{3,2}$ and $k_{4,2}$ from orbit 3 and 4 as examples, respectively. Satellite $k_{3,2}$ has the first visit after around 9 hours. However, satellite $k_{4,2}$ only visits the PS in the first 6 hours and after that does not have any visit. 
The beginning and duration of each visit between satellites and the PS are predictable. This is because the movement patterns of the satellite and PS in space are known in advance. To adapt the well-known FedAvg scheme introduced in \cite{mcmahan2017communication}, the non-availability between satellites and the PS must be taken into account. Hence, to enable the synchronous FL, we propose two schemes named FedNonISL and FedISL. FedNonISL implements the FedAvg by considering the visiting periods between PS and satellites. While FedISL uses the intra-plane ISLs and the predictability of satellites positions to schedule the exchanging model parameters between PS and satellites in each orbit.

%%%%%%%%%%%% FedNonISL scheme %%%%%%%%%%%%%%%%%
\subsection{Canonical FL for the Satellites: FedNonISL}

To implement the canonical FedAvg \nasrin{\cite{mcmahan2017communication}}, the model aggregation step needs the exchange of information between satellites and PS. As explained, the satellites and PS are not available all the time. In this regard, without the scheduling and using the intra-plane ISLs, the PS needs to wait for all the satellites to receive the model parameters.
To do this as Eq. \ref{eq: FL parameters}, all $K$ satellites must send the model parameters to the PS. 

Each satellite $k$ from any orbit $p$ receives the initial global parameters from the PS in its visit period. Then, it trains the model based on some local gradient steps as Eq. \ref{eq: GD} and derives the new local parameters. It transmits the updated parameters to the PS directly in its visibility period. For each global step, PS waits to receive all the model parameters from all $K$ satellites in their visit period. The time required for waiting depends on the constellation specifications and the position of PS. After receiving the parameters by the PS and aggregating them, the PS must again send the aggregated parameters to all LEOs to train. It requires the PS to wait for all LEOs to visit again. This process will continue for $N$ global updates. 

\subsection{Federated Learning with Intra-Plane ISLs: FedISL}
% The reason why we proposed this scheme
To cope with intermittent connectivity between satellites and the PS, we design a communication scheme, denoted as FedISL, that employs intra-plane ISLs and the predictability of satellite movements to enable synchronous FL. In particular, the satellites in each orbit build a ring network using ISLs, where each satellite can connect to both of its direct orbital neighbors. Communication between the PS and satellites within one orbit is routed through the satellite that has the current best connection to the PS. Such routing decisions are enabled by the predictability of orbital mechanics and require no additional communication between the satellites. The resulting topology is a star network of ring clusters that is not subject to the long delays involved in FedNonISL.

Each communication/computation round is split into three different phases: parameter distribution, computation, and aggregation. In the distribution phase, one satellite per orbit, denoted as the source, receives the model parameters from the PS and distributes them using ISLs to its neighboring satellites. These continue to propagate the parameters through their respective ISLs until all satellites have the current global model parameters. In particular, idle satellites from each orbit  \nasrin{that are not in parameter distribution, computation, and aggregation phases,} connect to the PS. Upon connection, the PS checks whether the connected satellite is from an orbit where the model parameters either are already received or are currently transmitted to. In the first case, the PS informs the satellite to wait for the parameters through ISL distribution. In the second case, the PS asks the satellite to reconnect after a short waiting time unless the parameters are received through ISL distribution in the meantime. If neither is the case, the PS transmits the global model and waits for the satellite to signal successful reception. Then, the PS marks the orbit as done and the satellite starts acting as the source of that orbit.

%Before distributing the model parameters, the source node uses predictive routing to determine the sink node, i.e., the node that is responsible for collecting and transmitting the updated model parameters to the PS.
Before distributing the model parameters, the source node uses predictive routing to determine the sink node that is responsible for collecting and transmitting the updated model parameters to the PS. \nasrin{This node is called sink.}
First, an estimate for the time needed to distribute the model parameters, complete the local computation tasks, and then collect the model updates from all satellites within orbit $p$ is computed as
\vspace{-1.5mm}
\begin{equation} \label{eq:aggregationtime}
T_{e,p} =  \Big\lfloor \frac{K_p}{2}  \Big \rfloor \cdot T_{c,p}+ T_{L,p} + \Big\lfloor \frac{K_p}{2}  \Big \rfloor \cdot T_{c,p},
\end{equation}
\vspace{-1.5mm}
where
\vspace{-1.2mm}
\begin{align}
T_{c,p} &=  \max_{k \in \mathcal K_p, i \in \mathcal N(k)}{t_c(k, i)}, &
T_{L,p} &= \max_{k \in \mathcal K_p} t_l(k)
\end{align}
are the maximum expected times required for learning and transmission between two adjacent satellites in orbit $p$, respectively, with $\mathcal N(k)$ being the neighbors of node $k$.

Then, the future position of all satellites within that orbit at time $t_{n} + T_{e,p}$ is computed using orbital mechanics, where $t_{n}$ is the current time. Among the satellites that will be within communication range of the PS, the satellite with the longest remaining contact time after $t_{n} + T_{e,p}$ is selected as the sink node. This routing decision is appended to the global model parameters, which are then propagated through both ISLs to the direct neighbors of the source. Then, the source node starts the computation phase. For all other satellites, upon reception of new model parameters via one of the ISLs, the receiving satellite forwards the model parameters to the next neighbor via its other ISL and then also advances into the next phase.

During the computation phase, the satellites update the model parameters with a local GD procedure as Eq.~(\ref{eq: GD}). Incoming ISL connections with global model parameters are dropped, i.e., the global model will only be forwarded once per communication/computation round. However, if a local model update is received from a neighboring node via ISL it is cached until completion of the computation phase.

Each satellite advances into the aggregation phase once its local GD procedure terminates. First, the routing graph $\mathcal G^n_p$ for the complete orbital plane $p$ is computed, where each satellite uses the route with the lowest number of hops to the sink node. If this decision is not unique, it selects the route via the neighboring satellite with the smallest ID. This results in a tree rooted at the sink node. If the satellite has children in this routing graph, it will wait until all parameter updates from those satellite are received and then computes the sum of the received parameters with its own parameters. In particular, at satellite $k$ and in epoch $n$, let $\mathcal I_k^n$ contain the parameter updates received via ISLs and $\pmb{w}_k^{n,I}$ be the result of the computation phase. Then, the outgoing partial aggregate is computed as
\vspace{-2mm}
\iffalse
\begin{equation} \label{eq:partialaggregate}
    \pmb{w}_{k,\mathrm{out}}^n = \frac{D_k}{\sum\nolimits_{k=1}^K D_k} \pmb{w}_k^{n,I} + \sum_{\pmb{w}\in\mathcal I_k^n} \pmb{w}.
\end{equation}
\fi
\begin{equation} \label{eq:partialaggregate}
    \nasrin{\pmb{w}_{k,\mathrm{out}}^n = {D_k} \pmb{w}_k^{n,I} + \sum\nolimits_{\pmb{w}\in\mathcal I_k^n} \pmb{w}.}
\end{equation}
%\vspace{-3mm}
%\begin{equation}
%    \pmb{w}_{k,\mathrm{out}}^n = \frac{D_k}{\sum_{k=1}^K D_k} \pmb{w}_k^{n,I} + %\sum_{\pmb{w}\in\mathcal I_k^n} \pmb{w}.
%\end{equation}

At all satellites except the sink, the partial aggregate $\pmb{w}_{k,\mathrm{out}}^n$ is then transmitted to the next satellite in the routing graph, i.e., to its parent vertex in this graph. Instead, the sink node sends $\pmb{w}_{k,\mathrm{out}}^n$ to the PS. After receiving all aggregates, PS derives $\bm{w}^{n+1}$ as \nasrin{$\pmb{w}^{n+1} = D^{-1} \sum\nolimits_{p=1}^{P} \pmb{w}_{p}^n$, }
%After receiving all aggregates, the PS derives $\bm{w}^n$ as
%\begin{equation} \label{eq:aggregateinPS}
%    \nasrin{\pmb{w}^{n+1} = D^{-1} \sum\nolimits_{p=1}^{P} \pmb{w}_{p}^n,}
%\end{equation}
where $\pmb{w}_p^n$ are the received partial aggregates of orbit $p$.
\nasrin{In case of a predictive routing failure, i.e., if satellites complete the process either too early or late, the sink node forwards the aggregated model parameters, again using predictive routing, until they reach a satellite that can communicate with the PS.} After successful transmission, each satellite moves into a new distribution phase.

\begin{algorithm}[t]
\caption{Parameter Server Operation}\label{alg:PS Procedure}
\small
\begin{algorithmic}[1] \color{black}
    \State \textbf{Initialize} epoch $n = 1$, model $\bm{w}^0$, number of orbital planes $P$%, orbital planes $\mathcal O_p$ for all $p$, where $\mathcal O_p$ contains satellites within orbit $p$
    \Statex \Comment Distribution Phase
     \vspace{-1mm}
    \State Set $\mathcal T^n = \emptyset$
    \While {$|\mathcal T^n| < P$}
    \State Upon incoming connection by satellite $k$: %Determine orbit $p$ of satellite $k$
    \State Determine orbit $p$ of satellite $k$
    \If {$p\notin\mathcal T^n$}
        \State Transmit $\bm{w}^{n-1}$ to satelite $k$
        \State Upon successful transmission, add $p$ to $\mathcal T^n$
    \Else \Comment Weights already sent to this orbit
        \State Terminate connection
    \EndIf
    \EndWhile
    \vspace{-1mm}
    \Statex \Comment Aggregation Phase
     \vspace{-1mm}
    \State Set $\mathcal R^n = \emptyset$, $\bm{w}^n = \bm{0}$
    \While {$|\mathcal R^n| < P$}
    \State Upon incoming connection by satellite $k$:
    \State Determine orbit $p$ of satellite $k$
    \If {$p\notin\mathcal R^n$ and received weights $\bm{w}_p^n$}
        \State Update $\bm{w}^n \gets \bm{w}^n  + \bm{w}_p^n$
        \State Add $p$ to $\mathcal R^n$
    \Else
        \State Terminate connection
    \EndIf
    \EndWhile
    \State Update ${w^n} \gets \frac{w^n}{D}$
\end{algorithmic}
\end{algorithm}

The PS, after it completes distribution, directly starts the aggregation phase and waits for incoming connections from the worker satellites. Once the partial aggregates from all orbits have been received, they are summed together to obtain the new version of the global model. This implements Eq.~\ref{eq: FL parameters} in a distributed manner. Then, the PS also advances to the next distribution phase. Incoming requests for a new global model are turned away during the aggregation phase. \bho{The described procedure is summarized in Algorithms~\ref{alg:PS Procedure} and~\ref{alg:Satellite Procedure} for the PS and and the satellites, respectively. Note that we have introduced two sets $\mathcal T^n, \mathcal R^n$ in Algorithm~\ref{alg:PS Procedure} to keep track of the states for each orbit, and a new variable $s_p^n$ in Algorithm~\ref{alg:Satellite Procedure} that denotes the sink node in orbit $p$ and epoch $n$. Also, for ease of exposition, Algorithms~\ref{alg:PS Procedure} only accepts connections sequentially. However, the extension to a parallel algorithm is quite straightforward.}

% Nasrin: In line 18 --> Add p to  $\mathcal R^n$
% In line 19 (outside of the if statement), Update $w^n$ based on the equation 14

\begin{algorithm}[t] \color{black}
\caption{Satellite Operation}\label{alg:Satellite Procedure}
\small
\begin{algorithmic}[1]
	\small
    \State \textbf{Initialize} epoch $n = 1$, satellite ID $k$ and orbital plane ID $p$
    \Loop \label{alg:sat:start}
    \Comment Parameter distribution phase
        \If {received $(s^n_p, \bm{w}^{n})$ from satellite $l$}
            \If {$s^n_p = -1$} \Comment Fallback routing
                \State Set $n \gets n - 1$, $\bm{w}_{k,\mathrm{out}}^n \gets \bm{w}^{n}$, Go to line~\ref{alg:sat:fallback}
                %\State Go to line~\ref{alg:sat:fallback}
            \EndIf
            \State Transmit $(s^n_p, \bm{w}^{n})$ to satellite $\mathcal N(k)\setminus\{l\}$, Go to line~\ref{alg:sat:learn}
            %\State Go to line~\ref{alg:sat:learn}
        \EndIf
        \If {PS became visible since last iteration or asked for reconnection}
            \State Connect to PS and ask for weights
            \If {received weights $\bm{w}^{n}$}
                \State Estimate aggregation time $T_{e,p}$ using \eqref{eq:aggregationtime}
                \State Set sink node $s_p^n$ to satellite with longest connection time to PS at current time $+ T_{e,p}$
                \State Transmit $(s^n_p, \bm{w}^{n})$ to neighbors $\mathcal N(k)$, Go to line~\ref{alg:sat:learn}
                %\State Go to line~\ref{alg:sat:learn}
            \EndIf
            \vspace{-1mm}
        \EndIf
    \EndLoop
    \vspace{-1.1mm}
    \Statex \Comment Computation phase
    \State $\bm{w}^{n,I}_k \gets \Call{Learning}{\bm{w}^{n}}$ \label{alg:sat:learn}
     \vspace{-1.5mm}
    \Statex\Comment Aggregation phase
    \State Build routing tree $\mathcal G^n_p$ to $s_p^n$ for whole orbital plane
    \State Initialize $\mathcal I_k^n \gets \emptyset$
    \If {Satellite $k$ has children in $\mathcal G^n_p$}
        \State Check receive buffer until all expected weights arrive
        \State Store weights in $\mathcal I_k^n$
    \EndIf
    \State Compute $\bm{w}_{k,\mathrm{out}}^n$ as in \eqref{eq:partialaggregate}
    \If {$s_p^n = k$}
        \If {PS is visible or will be in short time} \label{alg:sat:fallback}
            \State Connect to PS and transmit $\bm{w}_{p}^n$
            %$\bm{w}_{k,\mathrm{out}}^n$
        \Else
            \State Determine which satellite $l\in\mathcal N(k)$ is closest to PS
            \State Transmit $(-1, \bm{w}_{k,\mathrm{out}}^n)$ to $l$
        \EndIf
    \Else
        \State Transmit $\bm{w}_{k,\mathrm{out}}^n$ to parent in $\mathcal G^n_p$
    \EndIf
    \State Set $n \gets n + 1$ and goto line~\ref{alg:sat:start}
\end{algorithmic}
\end{algorithm}

\begin{figure}
\begin{center}
\includegraphics[width=0.4\textwidth]{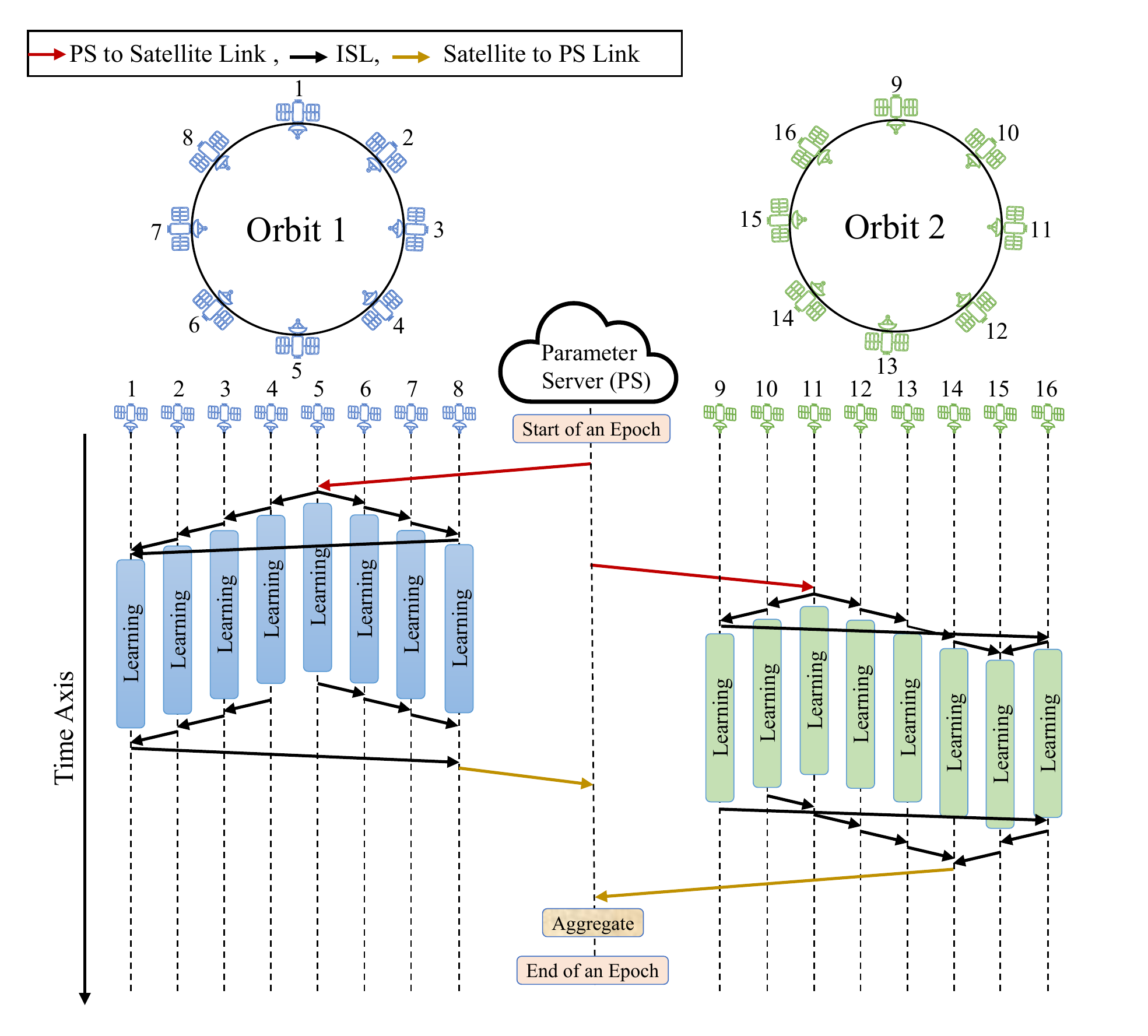}
\end{center}
\caption{Example for FedISL scheme with two orbits}
\vspace{-2.5ex}
\label{fig: FedISL scheme description}
\end{figure}
%\todo[inline]{Nasrin, please add an illustration of the routing tree for your example. Note that it is an undirected graph, so only edges, no arrows. Consider doing it in tikz for practice.}
% Example of the receiving the parameters from the server process  
%We illustrate this procedure with an example. Consider Fig.~\ref{fig: FedISL scheme description} with two orbits which we show with 1 and 2. 
We illustrate this procedure with an example in Fig.~\ref{fig: FedISL scheme description} with two orbits 1 and 2. Each orbit consists of 8 satellites. Satellites 1-8 and 9-16 are located in orbits 1 and 2, respectively. In orbit 1, the PS visits and transmits the model parameters to satellite 5. Then, this satellite determines the routing decision, appends it to the model parameters, then sends it to satellites 4 and 6 over the ISLs and starts the training. Satellites 4 and 6 relay the received packet to 3 and 7, respectively. Finally, satellite 1 stops the transmission to the other satellites because it receives the weights from both 2 and 8. 
%The routing tree for parameter distribution phase of orbit 1 is shown in \label{fig: Routing tree for parameter distribution phase}. 
In orbit 2, the PS transmits the model parameters to 11. Then, this source of orbit 2 derives the routing tree and transmits the model parameters to adjacent satellites and this process continues as orbit 1.
%\nasrin{Distributing of the model parameters in one orbit is described in Algorithm~\ref{alg:Distribute MP}, where $N^+(k)$ and $N^-(k)$ denote the nearest neighbor of satellite $k$ in the same and reverse directions of satellite movement, respectively.}  

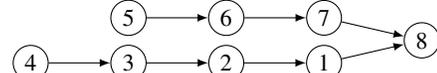
\begin{figure}[t]
\centering
\begin{tikzpicture}[grow=left, every node/.style={circle,draw,font=\small,inner sep=2pt}, edge from parent/.style={draw,latex-},sibling distance=6mm,level distance=13mm]
    \node {8}
        child {node {7} child {node {6} child {node{5}}}}
        child {node {1} child {node {2} child {node{3} child {node{4}}}}};
\end{tikzpicture}
\vspace{-2.5ex}
\caption{Routing tree for aggregation in Orbit 1 of Fig.~\ref{fig: FedISL scheme description} }
\label{fig: Routing tree for aggregation phase}
\end{figure}

As Fig. \ref{fig: FedISL scheme description} shows,  satellite 8 is selected as the sink satellite in orbit 1. Then, satellites 4 and 5 transmit the parameters to 3 and 6, respectively. In the next step, 3 and 6 aggregate their own parameters with the model parameters of 4 and 5. Because of the satellites' positions, after this step, 3 and 6 send the aggregated model parameters to 2 and 7 and they sum the received and own parameters. This process continues until satellite 1 transmits the aggregated parameters to satellite 8. This satellite aggregates the \nasrin{model parameters} received on both sides of ISLs and its own \nasrin{parameters}.
%This satellite aggregates the weights received on both sides of ISLs and its own weights.
The routing tree for aggregating of the model parameters in orbit 1 is depicted in \ref{fig: Routing tree for aggregation phase}. Finally, the sink sends the aggregated parameters to the PS. PS waits for the parameters from orbit 2. The process to aggregate the model parameters is done as orbit 1 with its routing tree. After receiving the model \nasrin{parameters} of orbit 2 by PS, the aggregation is done by PS. 
%After receiving the model weights of orbit 2 by PS, the aggregation is done by PS. 
%\nasrin{Aggregation of the model parameters in one orbit is expressed in Algorithm 2.}   

\iffalse
\begin{figure}[t]
\centering
\begin{tikzpicture}[grow=right, every node/.style={circle,draw}]
    \node {5}
        child {node {4} child {node {3} child {node{2} child {node{1}}}}}
        child {node {6} child {node {7} child {node{8} child {node{1}}}}}
        ;
\end{tikzpicture}
\caption{Routing tree for parameter distribution phase}
\label{fig: Routing tree for parameter distribution phase}
\end{figure}
\fi

\section{Experimental Results}

We present the convergence results of the proposed FedNonISL and FedISL schemes. We consider a walker constellation with a total of 40 LEO satellites distributed in 5 orbital planes in an altitude of 2000 km with inclination $\ang{80}$. We evaluate the performance of the proposed schemes with three different cases for the position of PS, one MEO satellite in altitude 20000 km with inclination $\ang{0}$, GS located in the North Pole, and Bremen, named as MEO, NP-GS, and BR-GS in the results, respectively. For NP-GS and BR-GS, the minimum elevation angle is set to $\ang{10}$.

Each satellite and the PS transmit the model parameters on a channel with a bandwidth of 20 MHz. We set the transmission power of each satellite and the PS to 40 \si{dBm} and the transmitter and receiver antenna gains to 6.98 \si{dBi}. We also set $f_c = 2.4 $ \si{GHz} and $T = 354.81$ \si{K}. We ignore the communication processing delays in transmitter and receiver because they are negligible compared to the sum of propagation and transmission delays.

For the ML task, a logistic regression model for image classification of handwritten digits 0-9 in MNIST data sets is considered \cite{li2018federated}. The learning rate is set to \nasrin{$\eta = 0.05$} and full batches are considered. We examine our proposed schemes with Non-IID data distributions. For the Non-IID data distribution, we assign 0-4 labels for 20 satellites and 5-9 to the other 20 satellites. We use the FedML research library \cite{he2020fedml}.

%We investigate the test accuracy of the considered model in terms of time for both schemes. For computation settings, $c_k  = 10^3$ and $f_k = 10^9$ are set \cite{zeng2020federated}.
We investigate the test accuracy of the considered model \nasrin{over} time for both schemes. For computation settings, $c_k  = 10^3$ and \nasrin{$\nu_k = 10^9 $} are set \cite{zeng2020federated}.
We show the performance of the FedISL and FedNonISL schemes for 4 days in Fig. \ref{test_acc_NIID}. 
%$f_k = 10^9$

 Fig. \ref{test_acc_NIID} shows that the FedISL scheme with the PS as NP-GS and MEO provides accuracy in the range of 83 \% after around 3:30 hours from the start of the simulation. With the comparison of the FedISL scheme with the MEO and NP-GS, in the beginning, satellites can receive the initial parameters from the MEO some minutes earlier than the NP-GS that resulting in higher accuracy in the first minutes. However, after some hours, their performance difference is negligible.
 
For the FedNonISL scheme with the MEO and NP-GS as the PSs, as Fig. \ref{test_acc_NIID} shows the time required for the test accuracy of 0.83$\%$ takes around 4 days with the MEO as the PS. \nasrin{Then, we have 29$\bm{\times}$ speed-up in the training process time for MEO as PS.} For the FedNonISL with the NP-GS, the accuracy is 0.78 $\%$. The reduction in the accuracy is the result of the intermittent connectivity between satellites and PSs which causes the PS to wait. We can see one MEO satellite as the PS can provide much faster convergence speed for the FedNonISL scheme compared with the NP-GS. \nasrin{This indicates to prefer MEOs to be used as PS rather the GS.}
%This shows the importance of MEO satellites as the PS.

%The other conclusion is that both schemes, FedISL and FedNonISL, with the PS in Bremen, Germany, can provide lower accuracy after 4 days. 
The other conclusion is that both schemes, FedISL and FedNonISL, with the PS in Bremen, Germany, \nasrin{shows} lower accuracy after 4 days. 
This is the result of the non-visibility of satellites and PS in combination with the synchronous learning scheme. The initial model parameters in the FedNonISL scheme can be received after 9:30 hours which makes a long delay in the convergence of FL. A better suited approach is the asynchronous algorithm proposed in \cite{razmi2021ground}.

We can conclude that implementing FL with the PS on Earth needs more investigation. However, with having one MEO satellite, this challenge can be solved. Using several GSs based on the considered constellation can be another solution to solve this issue. \nasrin{FedISL also reduces the traffic in the PS. For our settings, the PS can experience 8$\bm{\times}$ traffic reduction at the PS over the FedNonISL.}

\iffalse
\begin{figure}
    \centering
    \includegraphics[width=0.55\textwidth]{FedISL_NonIID_result.eps}
    \caption{Test-Accuracy for FedISL and FedNonISL schemes with non-IID data distribution}
    \label{test_acc_NIID}
\end{figure}
\fi

\begin{figure}
\begin{tikzpicture}
\begin{axis}[
yminorgrids = true, 
legend entries = {FedISL-MEO, FedNonISL-MEO, FedISL-NP-GS, FedNonISL-NP-GS, FedISL-BR-GS, FedNonISL-BR-GS},
xlabel={Time [h]},
ylabel={Test Accuracy},
grid=major,
grid=minor,
xtick = {0,24,...,96},
xmin = 0,
xmax = 96,
ymin = 0.05,
ymax = 0.9,
minor x tick num = 3,
minor y tick num = 1,
grid = major,
width=1*\axisdefaultwidth,
height=1*\axisdefaultheight,
legend pos=south east,
legend cell align=left,
legend style={font=\footnotesize}
]

\addplot[color=red,smooth,dashed] table [x=FedISL_MEO, y=accuracy_NIID_FedISL_MEO, col sep=comma] {CSV_FedISL_ICC.csv};

\addplot[color=red,smooth,thick] table [x=FedNonISL_MEO, y=accuracy_NIID_FedNonISL_MEO, col sep=comma] {CSV_FedISL_ICC.csv};
%color=g tik
\addplot[color=black,smooth,densely dashdotted] table [x=FedISL_NP_elevation_angle, y=accuracy_NIID_FedISL_elevation_NP, col sep=comma] {CSV_FedISL_ICC.csv};

\addplot[color=black,smooth,thick] table [x=FedNonISL_NP_elevation_angle, y=accuracy_NIID_FedNonISL_elevation_NP, col sep=comma] {CSV_FedISL_ICC.csv};

\addplot[color=blue,smooth,thick,dashed] table [x=FedISL_BR_elevation_angle, y=accuracy_NIID_FedISL_elevation_BR, col sep=comma] {CSV_FedISL_ICC.csv};
\addplot[color=blue,smooth,thick] table [x=FedNonISL_BR_elevation_angle, y=accuracy_NIID_FedNonISL_elevation_BR, col sep=comma] {CSV_FedISL_ICC.csv};

\end{axis}
\end{tikzpicture}
\vspace{-2.5ex}
\caption{Accuracy for PS as MEO, GS at North Pole and Bremen in 4 days.}
\label{test_acc_NIID}
\end{figure}
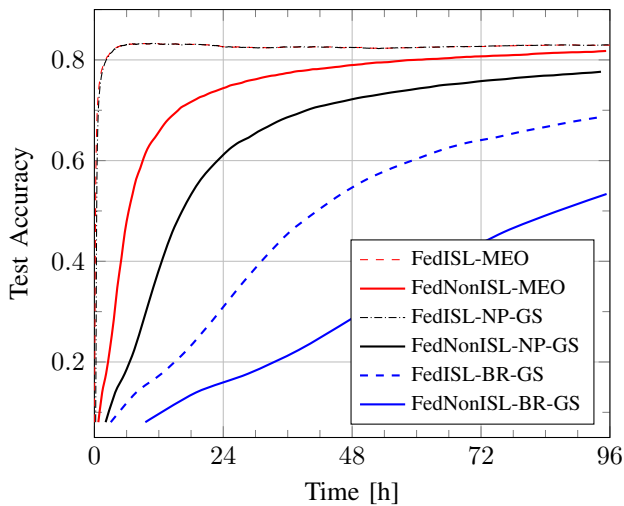

\section{Conclusion}
We have proposed a novel communication scheme to support FL in LEO satellite mega-constellations equipped with intra-plane ISLs\nasrin{, named as FedISL}. \nasrin{FedISL} relies on predictive scheduling and improves the convergence time of FedAvg significantly over the baseline. Owing to partial aggregation, this does not come at the cost of \nasrin{higher} communication cost. Indeed, while the baseline implementation, \nasrin{FedNonISL}, concentrates all communication at the PS, the proposed protocol evenly distributes the communication within the constellation resulting in only two PS data exchanges per orbital plane. \nasrin{The performance was evaluated numerically and highlights superiority of the proposed method.} \nasrin{Especially for a PS in MEO and also for a carefully located ground station, it shows promising performance.} However, the performance for suboptimally located ground stations shows still potential for improvement. Possible solution approaches involve careful worker scheduling, asynchronous algorithms (see, e.g., \cite{razmi2021ground}), and inter-orbit ISLs.
%I would split it up in to simple but clear sentences.
%1) ISL show promising gains
%2) to use MEOs as PS is promising 

%We have proposed a novel communication scheme to support FL in LEO satellite mega-constellations equipped with intra-plane ISLs. It relies on predictive scheduling and improves the convergence time of FedAvg significantly over the baseline. Owing to partial aggregation, this does not come at the cost of increased communication cost. Indeed, while the baseline implementation concentrates all communication at the PS, the proposed protocol evenly distributes the communication within the constellation resulting in only two PS data exchanges per orbital plane. The performance was evaluated numerically and highlights superiority of the proposed method, especially for a PS in MEO but also for a carefully located ground station. However, the performance for suboptimally located ground stations shows still potential for improvement. Possible solution approaches involve careful worker scheduling, asynchronous algorithms (see, e.g., \cite{razmi2021ground}), and inter-orbit ISLs.
%\balance

\bibliographystyle{IEEEtran}
\bibliography{IEEEtrancfg,IEEEabrv,references}

\begin{thebibliography}{10}
\providecommand{\url}[1]{#1}
\csname url@samestyle\endcsname
\providecommand{\newblock}{\relax}
\providecommand{\bibinfo}[2]{#2}
\providecommand{\BIBentrySTDinterwordspacing}{\spaceskip=0pt\relax}
\providecommand{\BIBentryALTinterwordstretchfactor}{4}
\providecommand{\BIBentryALTinterwordspacing}{\spaceskip=\fontdimen2\font plus
\BIBentryALTinterwordstretchfactor\fontdimen3\font minus
  \fontdimen4\font\relax}
\providecommand{\BIBforeignlanguage}[2]{{%
\expandafter\ifx\csname l@#1\endcsname\relax
\typeout{** WARNING: IEEEtran.bst: No hyphenation pattern has been}%
\typeout{** loaded for the language `#1'. Using the pattern for}%
\typeout{** the default language instead.}%
\else
\language=\csname l@#1\endcsname
\fi
#2}}
\providecommand{\BIBdecl}{\relax}
\BIBdecl

\bibitem{sweeting2018modern}
M.~Sweeting, ``Modern small satellites --- changing the economics of space,''
  \emph{Proc. {IEEE}}, vol. 106, no.~3, pp. 343--361, 2018.

\bibitem{9217520}
I.~Leyva-Mayorga, B.~Soret, M.~Röper, D.~Wübben, B.~Matthiesen, A.~Dekorsy,
  and P.~Popovski, ``{LEO} small-satellite constellations for {5G} and
  beyond-{5G} communications,'' \emph{{IEEE} Access}, vol.~8, 2020.

\bibitem{mateo2021towards}
G.~Mateo-Garcia, J.~Veitch-Michaelis, L.~Smith, S.~V. Oprea, G.~Schumann,
  Y.~Gal, A.~G. Baydin, and D.~Backes, ``Towards global flood mapping onboard
  low cost satellites with machine learning,'' \emph{Sci. Rep.}, vol.~11,
  no.~1, 2021.

\bibitem{da2005first}
A.~da~Silva~Curiel, L.~Boland, J.~Cooksley, M.~Bekhti, P.~Stephens, W.~Sun, and
  M.~Sweeting, ``First results from the disaster monitoring constellation
  ({DMC}),'' \emph{Acta Astronautica}, vol.~56, no. 1-2, 2005.

\bibitem{vazquez2021machine}
M.~{\'A}. V{\'a}zquez, P.~Henarejos, I.~Pappalardo, E.~Grechi, J.~Fort, J.~C.
  Gil, and R.~M. Lancellotti, ``Machine learning for satellite communications
  operations,'' \emph{{IEEE} Commun. Mag.}, vol.~59, no.~2, pp. 22--27, 2021.

\bibitem{giuffrida2020cloudscout}
G.~Giuffrida, L.~Diana, F.~de~Gioia, G.~Benelli, G.~Meoni, M.~Donati, and
  L.~Fanucci, ``Cloudscout: a deep neural network for on-board cloud detection
  on hyperspectral images,'' \emph{Remote Sensing}, vol.~12, no.~14, p. 2205,
  2020.

\bibitem{mcmahan2017communication}
H.~B. McMahan, E.~Moore, D.~Ramage, S.~Hampson, and B.~Agüera~y Arcas,
  ``Communication-efficient learning of deep networks from decentralized
  data,'' ser. Proc. Mach. Learn. Res. (PMLR), vol.~54, 2017.

\bibitem{razmi2021ground}
\BIBentryALTinterwordspacing
N.~Razmi, B.~Matthiesen, A.~Dekorsy, and P.~Popovski, ``Ground-assisted
  federated learning in {LEO} satellite constellations,'' Sep. 2021. [Online].
  Available: \url{https://arxiv.org/abs/2109.01348}
\BIBentrySTDinterwordspacing

\bibitem{9327501}
I.~Leyva-Mayorga, B.~Soret, and P.~Popovski, ``Inter-plane inter-satellite
  connectivity in dense {LEO} constellations,'' \emph{{IEEE} Trans. Wireless
  Commun.}, vol.~20, no.~6, pp. 3430--3443, 2021.

\bibitem{lee2020tornadoaggregate}
J.-W. Lee, J.~Oh, S.~Lim, S.-Y. Yun, and J.-G. Lee, ``Tornadoaggregate:
  Accurate and scalable federated learning via the ring-based architecture,''
  \emph{Conf. on Artificial Intell (AAAI).}, 2021.

\bibitem{duan2020self}
M.~Duan, D.~Liu, X.~Chen, R.~Liu, Y.~Tan, and L.~Liang, ``Self-balancing
  federated learning with global imbalanced data in mobile systems,''
  \emph{{IEEE} Trans. Parallel Distrib. Syst.}, vol.~32, no.~1, pp. 59--71,
  2020.

\bibitem{wang2021efficient}
Z.~Wang, Y.~Hu, J.~Xiao, and C.~Wu, ``Efficient ring-topology decentralized
  federated learning with deep generative models for industrial artificial
  intelligent,'' \emph{arXiv preprint arXiv:2104.08100}, 2021.

\bibitem{li2020unified}
Z.~Li and P.~Richt{\'a}rik, ``A unified analysis of stochastic gradient methods
  for nonconvex federated optimization,'' \emph{Neural Inf. Proc. Systems (
  NeurIPS)}, 2020.

\bibitem{nguyen2020toward}
M.~N. Nguyen, N.~H. Tran, Y.~K. Tun, Z.~Han, and C.~S. Hong, ``Toward multiple
  federated learning services resource sharing in mobile edge networks,''
  \emph{{IEEE} Trans. Mobile Comput.}, June 2021.

\bibitem{ippolito2017satellite}
L.~J. Ippolito~Jr, \emph{Satellite Communications Systems Engineering}.\hskip
  1em plus 0.5em minus 0.4em\relax John Wiley \& Sons, 2017.

\bibitem{li2018federated}
T.~Li, A.~K. Sahu, M.~Zaheer, M.~Sanjabi, A.~Talwalkar, and V.~Smith,
  ``Federated optimization in heterogeneous networks,'' \emph{arXiv preprint
  arXiv:1812.06127}, 2018.

\bibitem{he2020fedml}
C.~He, S.~Li, J.~So, X.~Zeng, M.~Zhang, H.~Wang, X.~Wang, P.~Vepakomma,
  A.~Singh, H.~Qiu \emph{et~al.}, ``Fedml: A research library and benchmark for
  federated machine learning,'' \emph{arXiv preprint arXiv:2007.13518}, 2020.

\bibitem{zeng2020federated}
T.~Zeng, O.~Semiari, M.~Mozaffari, M.~Chen, W.~Saad, and M.~Bennis, ``Federated
  learning in the sky: Joint power allocation and scheduling with uav swarms,''
  in \emph{IEEE Int. Conf. on Commun. (ICC)}, 2020, pp. 1--6.

\end{thebibliography}
\end{document}